\documentclass[aps,prd,twocolumn,showpacs,superscriptaddress,groupedaddress]{revtex4}  
\usepackage{graphicx}  
\usepackage{dcolumn}   
\usepackage{bm}        
\usepackage{amssymb}   

\usepackage{amsmath}
\usepackage{multirow}
\usepackage{comment}

\def\Etmiss{{E\kern-0.6em\slash}_{T}}

\begin{document}

\title{Measurement of the Electric Charge of the Top Quark in $\boldsymbol{t\bar{t}}$ Events} 
\affiliation{LAFEX, Centro Brasileiro de Pesquisas F\'{i}sicas, Rio de Janeiro, Brazil}
\affiliation{Universidade do Estado do Rio de Janeiro, Rio de Janeiro, Brazil}
\affiliation{Universidade Federal do ABC, Santo Andr\'e, Brazil}
\affiliation{University of Science and Technology of China, Hefei, People's Republic of China}
\affiliation{Universidad de los Andes, Bogot\'a, Colombia}
\affiliation{Charles University, Faculty of Mathematics and Physics, Center for Particle Physics, Prague, Czech Republic}
\affiliation{Czech Technical University in Prague, Prague, Czech Republic}
\affiliation{Institute of Physics, Academy of Sciences of the Czech Republic, Prague, Czech Republic}
\affiliation{Universidad San Francisco de Quito, Quito, Ecuador}
\affiliation{LPC, Universit\'e Blaise Pascal, CNRS/IN2P3, Clermont, France}
\affiliation{LPSC, Universit\'e Joseph Fourier Grenoble 1, CNRS/IN2P3, Institut National Polytechnique de Grenoble, Grenoble, France}
\affiliation{CPPM, Aix-Marseille Universit\'e, CNRS/IN2P3, Marseille, France}
\affiliation{LAL, Universit\'e Paris-Sud, CNRS/IN2P3, Orsay, France}
\affiliation{LPNHE, Universit\'es Paris VI and VII, CNRS/IN2P3, Paris, France}
\affiliation{CEA, Irfu, SPP, Saclay, France}
\affiliation{IPHC, Universit\'e de Strasbourg, CNRS/IN2P3, Strasbourg, France}
\affiliation{IPNL, Universit\'e Lyon 1, CNRS/IN2P3, Villeurbanne, France and Universit\'e de Lyon, Lyon, France}
\affiliation{III. Physikalisches Institut A, RWTH Aachen University, Aachen, Germany}
\affiliation{Physikalisches Institut, Universit\"at Freiburg, Freiburg, Germany}
\affiliation{II. Physikalisches Institut, Georg-August-Universit\"at G\"ottingen, G\"ottingen, Germany}
\affiliation{Institut f\"ur Physik, Universit\"at Mainz, Mainz, Germany}
\affiliation{Ludwig-Maximilians-Universit\"at M\"unchen, M\"unchen, Germany}
\affiliation{Panjab University, Chandigarh, India}
\affiliation{Delhi University, Delhi, India}
\affiliation{Tata Institute of Fundamental Research, Mumbai, India}
\affiliation{University College Dublin, Dublin, Ireland}
\affiliation{Korea Detector Laboratory, Korea University, Seoul, Korea}
\affiliation{CINVESTAV, Mexico City, Mexico}
\affiliation{Nikhef, Science Park, Amsterdam, the Netherlands}
\affiliation{Radboud University Nijmegen, Nijmegen, the Netherlands}
\affiliation{Joint Institute for Nuclear Research, Dubna, Russia}
\affiliation{Institute for Theoretical and Experimental Physics, Moscow, Russia}
\affiliation{Moscow State University, Moscow, Russia}
\affiliation{Institute for High Energy Physics, Protvino, Russia}
\affiliation{Petersburg Nuclear Physics Institute, St. Petersburg, Russia}
\affiliation{Instituci\'{o} Catalana de Recerca i Estudis Avan\c{c}ats (ICREA) and Institut de F\'{i}sica d'Altes Energies (IFAE), Barcelona, Spain}
\affiliation{Uppsala University, Uppsala, Sweden}
\affiliation{Taras Shevchenko National University of Kyiv, Kiev, Ukraine}
\affiliation{Lancaster University, Lancaster LA1 4YB, United Kingdom}
\affiliation{Imperial College London, London SW7 2AZ, United Kingdom}
\affiliation{The University of Manchester, Manchester M13 9PL, United Kingdom}
\affiliation{University of Arizona, Tucson, Arizona 85721, USA}
\affiliation{University of California Riverside, Riverside, California 92521, USA}
\affiliation{Florida State University, Tallahassee, Florida 32306, USA}
\affiliation{Fermi National Accelerator Laboratory, Batavia, Illinois 60510, USA}
\affiliation{University of Illinois at Chicago, Chicago, Illinois 60607, USA}
\affiliation{Northern Illinois University, DeKalb, Illinois 60115, USA}
\affiliation{Northwestern University, Evanston, Illinois 60208, USA}
\affiliation{Indiana University, Bloomington, Indiana 47405, USA}
\affiliation{Purdue University Calumet, Hammond, Indiana 46323, USA}
\affiliation{University of Notre Dame, Notre Dame, Indiana 46556, USA}
\affiliation{Iowa State University, Ames, Iowa 50011, USA}
\affiliation{University of Kansas, Lawrence, Kansas 66045, USA}
\affiliation{Louisiana Tech University, Ruston, Louisiana 71272, USA}
\affiliation{Northeastern University, Boston, Massachusetts 02115, USA}
\affiliation{University of Michigan, Ann Arbor, Michigan 48109, USA}
\affiliation{Michigan State University, East Lansing, Michigan 48824, USA}
\affiliation{University of Mississippi, University, Mississippi 38677, USA}
\affiliation{University of Nebraska, Lincoln, Nebraska 68588, USA}
\affiliation{Rutgers University, Piscataway, New Jersey 08855, USA}
\affiliation{Princeton University, Princeton, New Jersey 08544, USA}
\affiliation{State University of New York, Buffalo, New York 14260, USA}
\affiliation{University of Rochester, Rochester, New York 14627, USA}
\affiliation{State University of New York, Stony Brook, New York 11794, USA}
\affiliation{Brookhaven National Laboratory, Upton, New York 11973, USA}
\affiliation{Langston University, Langston, Oklahoma 73050, USA}
\affiliation{University of Oklahoma, Norman, Oklahoma 73019, USA}
\affiliation{Oklahoma State University, Stillwater, Oklahoma 74078, USA}
\affiliation{Brown University, Providence, Rhode Island 02912, USA}
\affiliation{University of Texas, Arlington, Texas 76019, USA}
\affiliation{Southern Methodist University, Dallas, Texas 75275, USA}
\affiliation{Rice University, Houston, Texas 77005, USA}
\affiliation{University of Virginia, Charlottesville, Virginia 22904, USA}
\affiliation{University of Washington, Seattle, Washington 98195, USA}
\author{V.M.~Abazov} \affiliation{Joint Institute for Nuclear Research, Dubna, Russia}
\author{B.~Abbott} \affiliation{University of Oklahoma, Norman, Oklahoma 73019, USA}
\author{B.S.~Acharya} \affiliation{Tata Institute of Fundamental Research, Mumbai, India}
\author{M.~Adams} \affiliation{University of Illinois at Chicago, Chicago, Illinois 60607, USA}
\author{T.~Adams} \affiliation{Florida State University, Tallahassee, Florida 32306, USA}
\author{J.P.~Agnew} \affiliation{The University of Manchester, Manchester M13 9PL, United Kingdom}
\author{G.D.~Alexeev} \affiliation{Joint Institute for Nuclear Research, Dubna, Russia}
\author{G.~Alkhazov} \affiliation{Petersburg Nuclear Physics Institute, St. Petersburg, Russia}
\author{A.~Alton$^{a}$} \affiliation{University of Michigan, Ann Arbor, Michigan 48109, USA}
\author{A.~Askew} \affiliation{Florida State University, Tallahassee, Florida 32306, USA}
% ...
%
%
\collaboration{The D0 Collaboration\footnote{with visitors from
$^{a}$Augustana College, Sioux Falls, SD, USA,
$^{b}$The University of Liverpool, Liverpool, UK,
$^{c}$DESY, Hamburg, Germany,
$^{d}$Universidad Michoacana de San Nicolas de Hidalgo, Morelia, Mexico
$^{e}$SLAC, Menlo Park, CA, USA,
$^{f}$University College London, London, UK,
$^{g}$Centro de Investigacion en Computacion - IPN, Mexico City, Mexico,
$^{h}$Universidade Estadual Paulista, S\~ao Paulo, Brazil,
$^{i}$Karlsruher Institut f\"ur Technologie (KIT) - Steinbuch Centre for Computing (SCC),
D-76128 Karlsruhe, Germany,
$^{j}$Office of Science, U.S. Department of Energy, Washington, D.C. 20585, USA,
$^{k}$American Association for the Advancement of Science, Washington, D.C. 20005, USA,
$^{l}$Kiev Institute for Nuclear Research, Kiev, Ukraine
and
$^{m}$University of Maryland, College Park, Maryland 20742, USA.
}} \noaffiliation
\vskip 0.25cm
\date{July 18, 2014}

\begin{abstract}
We present a measurement of the electric charge of top quarks using $t\bar{t}$ events 
produced in $p\bar{p}$ collisions at the Tevatron. 
The analysis is based on fully reconstructed $t\bar{t}$ pairs in lepton+jets final states.
Using data corresponding to 5.3 $\rm fb^{-1}$ of integrated luminosity,
we exclude the hypothesis that the top quark has a charge of $Q=-4/3\,e$ 
at a significance greater than 5 standard deviations.
We also place an upper limit of 0.46 on the fraction of such quarks that can be present 
in an admixture with the standard model top quarks ($Q=+2/3\,e$) at a 95\% confidence level. 
\end{abstract}

\maketitle

\newcommand{\dzero}     {D\O}
\newcommand{\smq}{$\rm{+2/3\,e}$}
\newcommand{\exq}{$\rm{-4/3\,e}$}
\newcommand{\ttb}{$t\bar{t}$}
\newcommand{\bq}{{\it b}-quark}
\newcommand{\qqb}{$q\bar{q}$}
\newcommand{\ie}{$\it{e}$}
\newcommand{\met}{\mbox{$\not\!\!E_T$}}
\newcommand{\tab}{\hspace*{0.5in}}
\newcommand{\up}{\vspace*{-1.0in}}
\newcommand{\ha}{\hspace{1mm}}
\newcommand\T{\rule{0pt}{2.25ex}}       


The top quark ($t$), discovered in $p\bar{p}$ collisions at the Tevatron in 1995~\cite{smTop},
fits within the standard model (SM) of particle physics as the companion of the $b$ quark 
in a weak-isospin doublet with an electric charge of $Q=+2/3\,e$.
\ttb~pairs via the strong interaction is the dominant production mode of top quarks at hadron colliders.
In the SM, the top quark decays $\approx 99.9\%$ of the time to a $W$ boson and a $b$ quark,
i.e., $t(+2/3\,e)\rightarrow W^{+}b$ and its charge conjugate. 
However, beyond the SM (BSM) a new quark with a charge of $Q=-4/3\thinspace e$ could contribute to the same final state
with the corresponding decay of $q_{\rm BSM}(-4/3\,e)\rightarrow W^{-}b$ and its charge conjugate~\cite{exot}\cite{bsm}.
This $q_{\rm BSM}$ is the down-type component of an exotic right-handed doublet 
with its companion quark having a charge of $Q=-1/3\thinspace e$~\cite{exot}.
The measured kinematic distributions of \ttb~events, in particular, 
the \ttb~mass spectrum, are consistent with the SM~\cite{diff_xsec}. 
This type of BSM quark would therefore be likely to appear in an admixture with SM top quarks in the \ttb~final state,
and evade detection unless the charge of the top quarks is measured explicitly.

Under an assumption that, except for the electric charge, 
all other properties of the BSM quark are identical to those of the SM top quark,
experimental limits have been placed on the BSM nature of the top quark 
in $p\bar{p}$ collisions at $\sqrt{s}=1.96$ TeV  by the D0 and CDF collaborations~\cite{prld0,prlcdf}, 
at 92\% and 99\% confidence levels, respectively.
A stringent exclusion has been reported by the ATLAS collaboration in $pp$ collisions at $\sqrt{s}=7$ TeV 
with a significance of more than 8 standard deviations (SD)~\cite{jhepatlas}.
In this paper, we discriminate between the SM top quark and the BSM quark under the above assumption,
using data accumulated with the D0 detector in $p\bar{p}$ collisions at $\sqrt{s}$=1.96 TeV 
corresponding to an integrated luminosity of 5.3 $\rm fb^{-1}$.
A kinematic fit to the \ttb~ final state~\cite{hitfit} is used to associate the $b$ jets with the $W$ candidates
and the charge of the $b$ jets is determined through a jet charge algorithm~\cite{qj_algo}.
We then extend the analysis to examine the additional possibility that the two types of quarks 
can contribute in an admixture to the top and antitop quarks of the $t\bar{t}$ final state
and place a stringent limit on the possible fraction of BSM quarks in the data.

The D0 detector~\cite{run2det} has a central tracking system, 
consisting of a silicon microstrip tracker and a central fiber tracker,
both located within a $1.9\,$T superconducting solenoidal magnet, 
optimized for tracking and vertexing at pseudorapidities $|\eta|<3$ and $|\eta|<2.5$, respectively~\cite{pseudorapidity}.
Central and forward preshower detectors are positioned just outside of the superconducting coil. 
A liquid-argon and uranium calorimeter has a central section covering pseudorapidities up to $|\eta|$ $\approx 1.1$, 
and two end sections that extend coverage to $|\eta|\approx 4.2$, with all three housed in separate cryostats~\cite{run1det}. 
An outer muon system for $|\eta|<2$ consists of a layer of tracking detectors and scintillation trigger counters 
in front of $1.8\,$T iron toroids, followed by two similar layers after the toroids~\cite{run2muon}.

We use the lepton+jets final states of $t\bar{t}$ candidate events,
where one $W$ boson decays leptonically ($W \rightarrow \ell \nu_{\ell}$ with $\ell$ denoting an electron ($e$) or a muon ($\mu$))
and the other into two light-flavor quarks ($W \rightarrow q'\bar{q}$).
The final state is therefore characterized by one isolated charged lepton of large
transverse momentum relative to the beam axis ($p_{T}$), four jets generally originating from the $q'$, $\bar{q}$, $b$ and $\bar{b}$ quarks,
and a significant imbalance in transverse momentum ({\mbox{$\not\!\!E_T$}}) resulting from the undetectable neutrino.

The event selection, object identification, and event simulation of signal and background follow the procedures described in Ref.~\cite{d0crossection}.
The primary interaction vertex (PV) from a $p\bar{p}$ collision must be reconstructed within 60$\,$cm of the detector center.
Electrons are required to have $p_T>20\,$GeV and $|\eta|<1.1$, and muons are required to have $p_T>20\,$GeV and $|\eta|<2.0$.
Electrons and muons from leptonic-tau decays ($W\rightarrow \tau\nu_{\tau}\rightarrow \ell\nu_{\ell}\nu_{\tau}$) are included in the analysis.
Jets are defined using an iterative cone algorithm~\cite{d0jets} with a radius 
$R=0.5$ in $(\eta,\phi)$ space, where $\phi$ is the azimuthal angle.
We select events with four or more jets with $p_T>20\,$GeV and $|\eta|<2.5$,
at least two of which are required to be identified (tagged) as $b$ jets through a neural network (NN) discriminant
at a threshold for which the tagging efficiency for $b$ jets is $\approx 55\%$ and the misidentification rate for light-flavor jets is $\approx 2\%$~\cite{btagging}.
The \met~is required to be greater than $20$ and $25\thinspace$GeV in the $e$+jets and $\mu$+jets events, respectively.

The production of $t\bar{t}$ pairs is simulated using the \textsc{alpgen} Monte Carlo (MC) generator~\cite{ALPGEN} with a top quark mass of $172.5\,$GeV. 
We use the \textsc{pythia}~\cite{PYTHIA} program for parton evolution and \textsc{geant}~\cite{GEANT} for simulating the D0 detector.
The dominant background process is $W+$jets production.
Several additional sources of background are also considered.
We simulate $W/Z+$jets, diboson ($WW, WZ,{~\rm{and}}~ZZ$), and single top quark productions using \textsc{alpgen}, \textsc{pythia}, and \textsc{comphep}~\cite{COMPHEP}, respectively.
The cross section for each background is normalized to next-to-leading-order predictions.
The contribution from multijet background is estimated from data using the ``matrix method''~\cite{matrix_method}.

The assignment of reconstructed objects to the products from $t\bar{t}$ decay
is achieved through a constrained kinematic fit~\cite{hitfit},
which, for each possible assignment, minimizes a $\chi^{2}$ function 
using the kinematic information of the reconstructed objects assuming the $t\bar{t}$ hypothesis for the final state objects.
As constraints for the fit, we use the conservation of energy and momentum and 
the masses of the $W$ boson and the top quark, $m_{W}=80.4\,$GeV and $m_{t}=172.5\,$GeV, respectively.
The $b$-tagged jets are assumed to be jets originating from $b$ quarks.
We utilize the {\mbox{$\not\!\!E_T$}} and the mass constraint on the leptonically decaying $W$ boson to infer the momentum of the neutrino.
The assignment with the lowest $\chi^{2}$ is used to reconstruct the \ttb~decay chain.
The $b$ jet that is paired with the $W$ boson that decays into two leptons or two quarks in the \ttb~event reconstruction is referred to, respectively, as $b_{\ell}$ or $b_{h}$.
The efficiency of correct assignment for $b_{\ell}$ and $b_{h}$ is $\approx 70\%$.

The charge of the lepton, $Q_{\ell}$, determines the charge of the leptonically decaying $W$ boson and 
consequently the opposite charge is assumed for the $W$ boson that decays to $q'\bar{q}$.
The charge of the quark initiating a jet is estimated from the reconstructed jet charge, $Q_{\rm j}$, 
using the method proposed in Ref.~\cite{qj_algo}.
The charge of the $b_{\ell}$ and $b_{h}$ is denoted as $Q_b^{\ell}$ and $Q_b^h$.
We combine the charges of each $W$ boson and its associated $b$ jet to compute the charge of the top quark
$Q^{\ell}_{t} = |Q_{\ell} + Q_b^{\ell}|$ for the top quark whose $W$ boson decays leptonically,
or $Q^{h}_{t} = |-Q_{\ell} + Q_b^h|$ for the top quark whose $W$ boson decays into quarks.
Using the modulus provides two quantities with the same distributions and thus a statistical benefit from merging them.
The values of the $b$-jet charges $Q_b^{\ell,h}$ are computed from a jet-charge algorithm
$Q_{\rm j} = {(\Sigma_{i}Q_{i} \cdot (p_{Ti})^{0.5})/}{(\Sigma_{i}(p_{Ti})^{0.5})}$,
where $i$ runs over all reconstructed tracks within the jet with the requirements that
each track has (i) a distance of closest approach within $0.2\,$cm relative to the PV and $p_{T}>0.5\thinspace$GeV,
and (ii) angular distance with respect to the jet axis $\Delta\it{R}(\rm track, jet)\equiv\sqrt{(\Delta\eta)^2+(\Delta\phi)^2}<0.5$,
with (iii) at least two tracks satisfying the above requirements.
These track criteria and the exponent of 0.5 are the results of an optimization of the algorithm using simulated \ttb~events.
Events with both $b_{\ell}$ and $b_h$ passing these additional tracking requirements are considered for further analysis. 
The corresponding efficiency is greater than 0.99.
Table~\ref{tab:yield} summarizes the sample composition and event yields, 
following the application of all selection criteria and reconstruction of the charge of the top quark.

\begin{table}[h]
  \begin{center}
    \begin{ruledtabular}
      \caption{\label{tab:yield}Sample composition and event yields following the implementation of all final selections.
        The quoted uncertainties include the statistical and systematic components. 
        The ``Dilepton $t\bar{t}$'' process represents $t\bar{t}$ events where both $W$ bosons decaying leptonically
        and ``Other'' includes the diboson and multijet processes. 
        The cross section $\sigma_{t\bar{t}}=7.24$~pb is used for $t\bar{t}$ events~\cite{nlo_xsec}.}
      \begin{tabular}{c l c l c}
        &Process && Expected events& \\
        \hline
        &$t\bar{t}$    &&~~~$263.1~_{-~18.7}^{+~17.9}$\,&  \T \\
        &Dilepton $t\bar{t}$ &&~~~~~~$9.4~_{-~0.7}^{+~0.6}$&  \T  \\
        &$W+\rm{jets}$ &&~~~~\,$12.7 \pm 2.1$&  \T  \\
        &$Z+\rm{jets}$ &&~~~~~~$1.4 \pm 0.5$&  \T  \\
        &Single top    &&~~~~~~$3.0 \pm 0.4$&  \T  \\
        &Other         &&~~~~~~$0.6 \pm 1.3$&  \T  \\
        \hline
        &Total expectation && ~~~$290.2~_{-~18.8}^{+~18.1}$\,&   \T \\[0.1em]
        \hline
        &Observed      &&~~~~~~~~~286&   \T \\[-0.2em]
      \end{tabular}
    \end{ruledtabular}
  \end{center}
\end{table}

The reconstruction of the jet charge is studied using a ``tag-and-probe'' method 
in an inclusive two-jet (dijet) data sample~\cite{4.3/fb} enriched in $b\bar{b}$ events, referred to as the ``tight dijet (TD) sample''.
The TD sample consists of events with:
(i) exactly two jets, each $b$ tagged, with $p_{T}>20\,$GeV and $|\eta|<2.5$;
(ii) the $\Delta{\phi}$ between the two jets of $>$ 3.0 radians; and
(iii) one jet (referred to as the ``tag jet'') containing a muon (referred to as the ``tagging muon'') 
with $p_{T}>4\,$GeV and $\Delta\it{R}$($\mu$, jet) $<$ 0.5.
We refer to the other jet in the dijet event as the ``probe jet''.

The TD sample contains a small fraction of $c\bar{c}$ and light parton dijet events.
The contribution from light partons is considered negligible
since a MC study finds the $b$-tagging efficiency for a light parton jet to be a factor of 20 smaller than for a $c$ jet.
The fraction of $c\bar{c}$ events in the TD sample is
estimated using the $p_{T}$ of the tagging muon
relative to the axis of the tag jet ($\it{p^{\rm rel}_T}$).
Muons originating from $b$-quark decays tend to have larger values in the $\it{p^{\rm rel}_T}$ spectrum than those from $c$-quark decays. 
We fit the $\it{p^{\rm rel}_T}$ distribution in data with the distributions 
from $b\bar{b}$ and $c\bar{c}$ events simulated using \textsc{pythia}
and find that the fraction of $c\bar{c}$ events in the TD sample is $x_c=0.093~\pm~0.009\thinspace(\rm stat)$.

The tagging muon in a dijet event is used to infer the charge of the quark initiating the tag jet
and consequently to determine whether the probe jet is initiated by a quark or an antiquark.
However, the tagging muon can originate from either a direct decay of a $B$ hadron or ``charge-flipping'' processes 
such as cascade decays of $B$ hadrons, e.g., $b \rightarrow c \rightarrow \ell$, or neutral $B$ meson mixings.
In the charge-flipping processes, the charge of the tagging muon can be opposite to that expected,
and therefore mistag the probe jet.
We simulate the charge-flipping processes using 
\mbox{$Z\rightarrow b\bar{b}$} (MC) events generated with the \textsc{pythia} program,
and find that a fraction $x_{f} = 0.352~\pm~0.008\thinspace (\rm stat)$ of the tagging muons 
have a charge opposite to that of the initial $b$ quarks~\cite{x_f}.
This value is verified by examining the charge correlation between muons in a subset of the TD data sample
 where an additional muon, having the same quality as the tagging muon, is required in the probe jet. 

The performance of the jet-charge algorithm depends on the kinematic properties of the jet,
mainly due to a dependence of the tracking efficiency on $p_T$ and $|\eta|$.
The kinematics of the dijet samples used to extract the jet-charge distributions differ from those of \ttb~events,
whose jet charges we wish to model.
To account for the differences in the performance arising from these kinematic differences,
we first re-weight the \ttb~MC events to get the same jet $p_{T}$ and $|\eta|$ spectra as observed in the dijet events. 
The ratio of the distributions of jet charge $Q_{\rm j}$ between the nominal and the re-weighted \ttb~samples is parametrized and used as a correction function.
This kinematic correction, 8\% on average, is applied to the charge distributions of the probe jets,
thereby modifying the jet-charge distributions from dijet data so that they model jets in \ttb~events.

\begin{figure}[h]
  \includegraphics[scale=0.40]{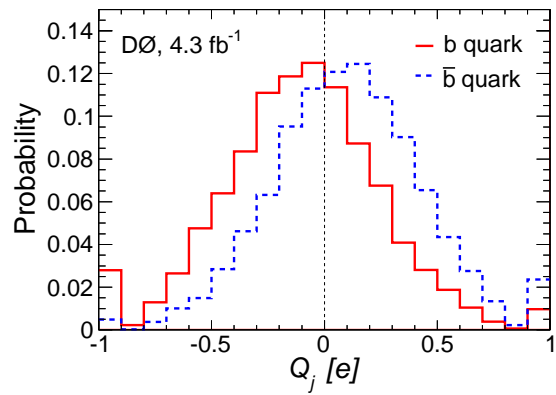}
  \caption{\label{fig:jet_charge_templates} 
    Distributions of charge templates for $b$ and $\bar{b}$ jets extracted from dijet events
    following the application of kinematic corrections described in the text.
  }
\end{figure}

To find the distributions of the jet charge for jets originating from $b$, $\bar{b}$, $c$, or $\bar{c}$ quarks, 
denoted as $\mathcal P_b(Q_{\rm j})$, $\mathcal P_{\bar{b}}(Q_{\rm j})$, $\mathcal P_c(Q_{\rm j})$, and $\mathcal P_{\bar{c}}(Q_{\rm j})$, respectively,
we utilize the distributions of the jet charge in probe jets of positive ($\mathcal P_+(Q_{\rm j})$) and negative ($\mathcal P_-(Q_{\rm j})$) tagging muons.
In the presence of $c\bar{c}$ contamination and of charge flipping processes (i.e., $x_c>0$ and $x_f>0$), we find

\begin{equation}
\label{eq:charge1}
\mathcal P_+(Q_{\rm j}) = (1- x_c)[x_f \mathcal P_{\bar{b}}(Q_{\rm j}) + (1- x_f) \mathcal P_{b}(Q_{\rm j})] + x_c \mathcal P_{\bar{c}}(Q_{\rm j}),
\end{equation}
and a similar expression for its charge-conjugate.
This requires extra inputs to solve for four unknown distributions. 
We use an additional data sample with a different composition from the TD sample.
This ``loose dijet (LD) sample'' is defined using the same selection criteria as for the TD events
except that the tag jets are not required to pass $b$-tagging requirements.
We find that the LD sample has a larger fraction of $c\bar{c}$ contributing with $\it{x_c'}$ = 0.352~$\pm$~0.014,
and a charge-flipping probability consistent with that found in the TD sample ($x_f'\simeq x_f$). 
The distributions for jet charge $\mathcal P'_{\pm}(Q_{\rm j})$ obtained from the probe jets in the LD sample provide the additional equations

\begin{equation}
\label{eq:charge2}
\mathcal P'_+(Q_{\rm j}) = (1-x'_c)[x_f \mathcal P_{\bar{b}}(Q_{\rm j}) + (1- x'_f)\mathcal P_{b}(Q_{\rm j})] + x_c' \mathcal P_{\bar{c}}(Q_{\rm j}), 
\end{equation}
and similarly for its charge conjugate.
The distributions for jet charge are constructed by solving Eqs.~(\ref{eq:charge1}) and (\ref{eq:charge2}), 
and their charge conjugate equations.
These charge templates $\mathcal P_b(Q_{\rm j})$, $\mathcal P_{\bar{b}}(Q_{\rm j})$, $\mathcal P_c(Q_{\rm j})$, and $\mathcal P_{\bar{c}}(Q_{\rm j})$, normalized to unity, 
serve as the probability density functions (PDF) for the charge of the jet originating from a given quark. 
The $b$-jet and $\bar{b}$-jet templates are shown in Fig.~\ref{fig:jet_charge_templates}.
The equivalent jet charge templates for each background process are derived through the same procedure as used for signal templates.

\begin{figure}[b]
  \includegraphics[scale=0.40]{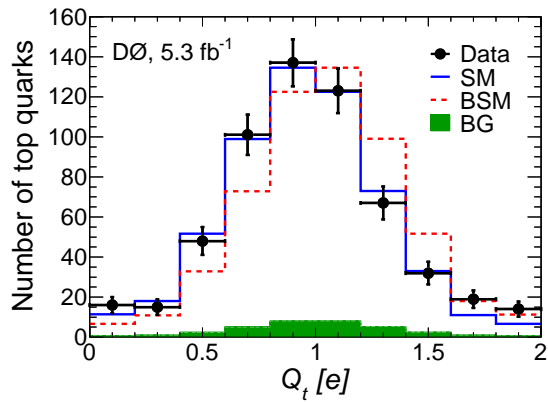}
  \caption{\label{fig:topcharge} 
    Combined distribution in the charge $Q_t$ for $t\bar{t}$ candidates in data 
    compared with expectations from the SM and the BSM. 
    The background contribution (BG) is represented by the green-shaded histogram. 
    The expected distributions are normalized to unity and used as the PDF
    $\mathcal P^{\rm SM}(Q_t)$ and $\mathcal P^{\rm BSM}(Q_t)$ in Eq.~(\ref{eq:lambda}).}
\end{figure}

The templates for jet charge are used to extract the top quark charges for each event, as follows. 
For the templates of SM top quark with $|Q|= 2/3\,e$, 
we obtain the charge observables $Q^{\ell}_t$ = $|Q_{\ell}$ + $Q_b^{\ell}|$ and $Q_t^h$ = $|-Q_{\ell}$ + $Q_b^h|$, 
while, for the BSM quark with $|Q|= 4/3\,e$, 
we obtain $Q_t^{\ell}$ = $|-Q_{\ell}$ + $Q_b^{\ell}|$ and $Q^h_t$ = $|Q_{\ell}$ + $Q_b^h|$.
We find the distributions of these two observables to be consistent and 
the correlation coefficient between them to be negligible ($\approx 4\%$)~\cite{Qt_corr}.
The 286 selected events in the lepton+jets final states provide 572 measurements of the top quark charge.
Figure~\ref{fig:topcharge} shows the combined distribution $Q_t$ ($Q^{\ell}_{t}$ and $Q^h_{t}$) observed in data, 
compared with the distributions expected for SM and BSM top quarks, including the background contribution.
For background events, no correlation is observed between the charge of the lepton and the $b$-jet assignment
and these combined observables contribute thereby equally to both distributions.

To measure the charge of the top quark,
we discriminate between the SM and BSM possibilities using a likelihood ratio:

\begin{equation}
\label{eq:lambda}
\Lambda=[\Pi_{i}\mathcal P^{\rm SM}(Q_t^i)] /[\Pi_{i}\mathcal P^{\rm BSM}(Q_t^i)],
\end{equation}
where $\mathcal P^{\rm SM}(Q_t^i)$ and $\mathcal P^{\rm BSM}(Q_t^i)$ are the probabilities of observing the top quark charge $Q_t^i$ 
under the SM and BSM hypotheses, respectively, according to the charge templates in Fig.~\ref{fig:topcharge}.
The superscript $i$ runs over all the 572 available measurements of $Q_t$.
The values of $\Lambda$ for the SM and BSM top quarks are evaluated through pseudo-experiments (PE)
using $\mathcal P^{\rm SM}(Q_t)$ and $\mathcal P^{\rm BSM}(Q_t)$, respectively.
A single PE consists of the same number of measurements as in data, 
randomly selected from the signal and background $Q_t$ distributions
according to the the sample composition in Table~\ref{tab:yield}.
Systematic uncertainties, detailed below, are accounted for in each PE by modifying the top quark charge templates as

\begin{equation}
\label{eq:2}
\mathcal P(Q_t) = \mathcal P^0(Q_t) + \sum_i \nu_i (\mathcal P^{i\pm}(Q_t) - \mathcal P^0(Q_t)),
\end{equation}
where $\mathcal P^{0}(Q_t)$ is the nominal probability distribution of $Q_t$, 
$\mathcal P^{i\pm}(Q_t)$ are those obtained from changes of $\pm 1$ SD made for systematic source $i$, 
and $\nu_i$ are nuisance parameters.
The $\nu_i$ are assumed to be uninteresting physical parameters, e.g., uncertainties that can be integrated over, and correspond to random variables drawn from a standard normal distribution.
We verify that variations in templates are linear with changes in the nuisance parameters.

\begin{figure}[t]
  \includegraphics[scale=0.40]{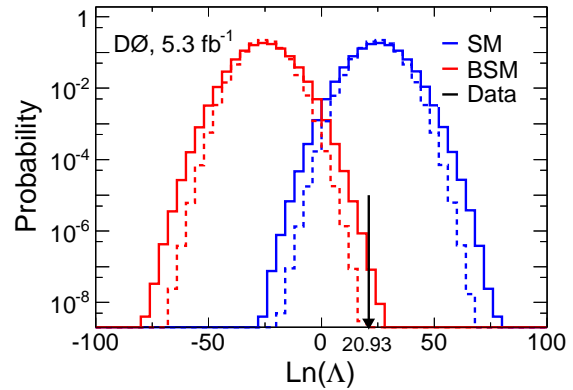}
  \caption{
    Distributions of ln($\Lambda$) for the SM (histograms on the right) and BSM (histograms on the left) models from $10^9$ PE, 
    compared to the measurement (arrow). 
    The solid lines show the expected distributions, 
    while the dashed histograms show the distributions expected 
    in the absence of systematic uncertainties.
    The value of ln($\Lambda_{\rm D}$) is displayed by the black vertical line.
  }
  \label{fig:llr_test}
\end{figure}

The data yields the value ln($\Lambda_{\rm D})=20.93$.
This value is compared to the distributions of $\Lambda$ for the SM and BSM assumptions shown in Fig.~\ref{fig:llr_test}. 
We find the measured $\Lambda_{\rm D}$ is consistent with the SM hypothesis 
and obtain a $p$-value of $6.0\times 10^{-8}$ under the BSM hypothesis, 
which corresponds to an exclusion of the BSM nature of the top quark with a significance of 5.4 SD.

We also consider the possibility that the observed distribution of events corresponds to a mixture of the SM and BSM top quarks. 
The fraction $f$ of the SM top quarks is determined using a binned likelihood fit.
The likelihood of the charge distribution in data is consistent with 
the sum of the SM and BSM templates that includes the background from Fig.~\ref{fig:topcharge},
with the number of events as a function of $Q_t$ given by:

\begin{equation}
n_i = f\times N\times \mathcal P_i^{\rm SM}(Q_t) + (1-f)\times N\times \mathcal P_i^{\rm BSM}(Q_t),
\end{equation}
where $N$ is the total number of measurements of $Q_t$, and $\mathcal P_i^{\rm SM}$ and $\mathcal P_i^{\rm BSM}$ 
is the probability of observing the SM and BSM top quarks, respectively, in bin $i$. 
The fraction $f$ is extracted by maximizing the likelihood without constraining $f$ to physically allowed values.

The systematic uncertainties on the fraction $f$ are listed in Table~\ref{tab:systematic},
and are classified in three categories: 
uncertainties related to (i) modeling of signal and background events; 
(ii) simulation of detector response; and (iii) analysis procedures and methods.
The maximum likelihood fit is repeated for each systematic source 
using the templates modified by the systematic effect, and
the resulting deviation from the nominal value is taken as the corresponding systematic uncertainty.

\begin{table}[h]
  \begin{center}
    \begin{ruledtabular}
      \caption{\label{tab:systematic} Summary of systematic uncertainties on the fraction $f$ of SM top quarks. The uncertainties are given in units of absolute value.}
      \begin{tabular}{l c c}
        Category & Source & Uncertainty \\
        \hline
        \multirow{5}{4.5em}{Signal/ background}
        &Signal modeling&$ 0.03$  \T \\
        &Initial/final state radiation&$ 0.01$  \T \\
        &Top quark mass&$ 0.01$  \T \\
        &Color reconnection&$ 0.01$  \T \\
        &Background normalization&$ 0.02$  \T \\
        \hline
        \multirow{7}{*}{Detector}
        &Lepton charge&\multirow{2}{*}{$ 0.01$}  \T \\[-0.1em]
        &mismeasurement& \\
        &Jet energy scale&$< 0.01$  \T \\
        &Jet identification&$ 0.02$  \T \\
        &Jet energy resolution&$ 0.02$  \T \\
        &$b$-tagging efficiency&$ 0.01$  \T \\
        &Luminosity&$ < 0.01$  \T \\
        \hline
        \multirow{6}{*}{Method}
        &$\Delta\phi$ in TD sample selection&$ 0.03$  \T \\
        &Determination of $x_c$&$ 0.01$  \T \\
        &Determination of $x_f$&$ 0.03$  \T \\
        &Kinematic corrections&$ 0.03$  \T \\
        &Dijet sample statistics&$ 0.07$  \T \\
        &MC template statistics&$ 0.03$  \T \\
        \hline
        \multicolumn{2}{l}{Total systematic uncertainty}& $ 0.11$  \T \\
      \end{tabular}
    \end{ruledtabular}
  \end{center}
\end{table}

The largest uncertainty, of 0.07, is due to the limited size of the selected dijet samples
used to model the charge templates for $b$-quark jets. 
Several systematic sources yield uncertainties on the measurement at the $\approx3\%$ level, such as 
(i) the determination of $x_f$, reflecting differences in the mixing parameters and decay rates of $B$ hadrons 
between the simulation and their latest experimental values~\cite{PDG}, 
(ii) the parametrization of the corrections for kinematic differences
in the distributions of jet charge for the dijet and \ttb~samples, 
and (iii) modeling of signal, where the effects of higher-order corrections, parton evolution, and hadronization 
are estimated using \ttb~events simulated with \textsc{mc@nlo}~\cite{MCNLO} interfaced 
with \textsc{herwig}~\cite{HERWIG} for parton evolution.

\begin{figure}[t]
  \includegraphics[scale=0.425]{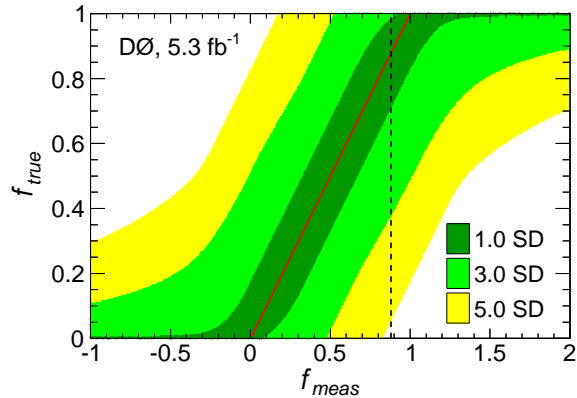}
  \caption{(color online) Confidence belts from the Feldman-Cousins approach for 1 SD (dark green), 3 SD (light green), and 5 SD (yellow). 
    The red solid line shows the average of the measured values $f_{\rm meas}$ for each input fraction $f_{\rm true}$ and the vertical dashed line represents the fraction ($f = 0.88$) observed in the data.
  }
  \label{fig:f-c} 
\end{figure}

The maximum likelihood fit to the top quark charge distribution in data 
yields the fraction $f = 0.88 \pm 0.13\thinspace(\rm {stat}) \pm 0.11\thinspace (\rm {syst})$.
We employ the ordering-principle suggested by Feldman and Cousins~\cite{f-c} to set limits on $f$.
The total uncertainty, i.e., the quadratic sum of the statistical and systematic uncertainties in Table~\ref{tab:systematic}, 
is assumed to be a Gaussian distribution in $f$. 
For the observed value, we find that the hypothesis that 
all top quarks in the data are BSM quarks is excluded at greater than 5 SD, as shown in Fig.~\ref{fig:f-c}, 
which is consistent with the results obtained from the likelihood ratio.
We also find a lower limit of $f=0.54$ at a 95\% CL, 
which corresponds to an equivalent upper limit on the fraction of BSM quarks of $f\leq$ 0.46 at the same level of significance.

In summary, using $b$-tagged jets in lepton+jets $t\bar{t}$ events in 5.3 $\rm{fb}^{-1}$ of $p\bar{p}$ data, 
we test the hypothesis that the particle assumed to be the SM top quark has an electric charge of $-4/3\,e$.
We exclude the possibility that all observed top quarks are BSM quarks at the level of more than 5 SD.
We also consider a possible admixture of such quarks with the SM top quarks 
and place an upper limit of 0.46 on the fraction of BSM quarks at a 95\% CL. 
The observed charge of the top quarks is in good agreement with the standard model.

%
We thank the staffs at Fermilab and collaborating institutions,
and acknowledge support from the
DOE and NSF (USA);
CEA and CNRS/IN2P3 (France);
MON, NRC KI and RFBR (Russia);
CNPq, FAPERJ, FAPESP and FUNDUNESP (Brazil);
DAE and DST (India);
Colciencias (Colombia);
CONACyT (Mexico);
NRF (Korea);
FOM (The Netherlands);
STFC and the Royal Society (United Kingdom);
MSMT and GACR (Czech Republic);
BMBF and DFG (Germany);
SFI (Ireland);
The Swedish Research Council (Sweden);
and
CAS and CNSF (China).
%



\begin{thebibliography}{99}
\bibitem{pseudorapidity} The pseudorapidity is defined as $\eta = - \ln[\tan(\theta/2)]$, where $\theta$ is the polar angle with respect to the beam axis measured at the center of the detector.
\bibitem{x_f} Using such simulations, we find that the charge-flipping fraction has contributions of 0.23 from cascade decays, 0.11 from neutral $B$ meson mixings, and 0.01 from misidentified electric charge.
\bibitem{Qt_corr} The two observables are expected to have the same distributions. However, two tagged jets are statistically independent and thus no strong correlation is expected.
\end{thebibliography}
\end{document}